\documentstyle[buckow]{article}

\newcommand{\be}{\begin{equation}}
\newcommand{\ee}{\end{equation}}

\newcommand{\bea}{\begin{eqnarray}}
\newcommand{\eea}{\end{eqnarray}}

\newcommand{\no}{\noindent}

\newcommand{\half}{\frac{1}{2}}

\def\NPB{{Nucl. Phys.} B}
\def\PLB{{Phys. Lett.} B}
\def\PRL{Phys. Rev. Lett.}

\def\PRD{{Phys. Rev.} D}

\begin {document}
\def\titleline{Topological Properties of the QCD Vacuum  
at $T = 0$ and $T \sim T_c$}
\def\authors{
Philippe de Forcrand \1ad , Margarita Garc\'{\i}a P\'erez \2ad,
James E.  Hetrick \3ad , Ion-Olimpiu Stamatescu \4ad
}
\def\addresses{
\1ad
SCSC, ETH-Z\"urich, CH-8092 Z\"urich, Switzerland\\
\2ad
Theory Division, CERN,
CH-1211, Geneve 23, Switzerland\\
\3ad
Physics Dept., University of the Pacific, Stockton, CA 95211-0197, USA\\
\4ad
FESt, Schmeilweg 5,
D-69118, Heidelberg\\
and\\
Institut f\"ur Theoretische Physik, Philosophenweg 16,
D-69120, Heidelberg, Germany
}

\def\abstracttext{
We study on the lattice the topology of $SU(2)$
and $SU(3)$ Yang-Mills theories at zero temperature and of $QCD$
at temperatures around the phase transition.
To smooth out dislocations and the UV noise we cool the configurations with
an action which has scale invariant instanton solutions for instanton size above
 $\sim 2.3$ lattice spacings. The corresponding ``improved" topological charge
stabilizes at an integer value after few cooling sweeps. At zero temperature
the susceptibility
calculated from this charge (about $(195\,{\rm MeV})^4$ for $SU(2)$ and 
$(185\,{\rm MeV})^4$ for $SU(3)$) agrees very well with the 
phenomenological expectation.
At the minimal amount 
of cooling necessary to resolve the structure in terms of instantons and 
anti-instantons we observe a dense ensemble where the total number of peaks 
is by a factor 5-10 larger than the net charge. The average size observed 
for these peaks at zero temperature 
is about 0.4-0.45 fm for $SU(2)$ and 0.5-0.6 fm for $SU(3)$.
The size distribution  changes 
very little with  further cooling, although in this process up to 90$\%$
of the peaks disappear by pair annihilation. For $QCD$ we observe below 
$T_c$ a drastic reduction of the topological susceptibility as an effect of 
the dynamical fermions. Nevertheless also here the instantons form a dense 
ensemble with general characteristics similar to those of the 
quenched theory. A further drop in the susceptibility above $T_c$ is also in
rough agreement with what has been observed for pure $SU(3)$. We see no clear signal for dominant formation of instanton - anti-instanton molecules.}

\makefront
\section{Introduction}

The vacuum of the 4-dimensional $SU(N_c)$ gauge theories  
has a non-trivial topological structure generated by different
coverings of the spatial sphere at $\infty$ with gauge transformations
from the $SU(2)$ subgroups. The tunneling
field configurations between these different vacua, the instantons,
 are self-dual  solutions of the classical, 
euclidean equations of motion of finite euclidean action and integer 
topological charge:
$S_0 = \int d^4x s(x) = 8 \pi^2,\ \ Q_0 = \int d^4x q(x) = \pm 1$. 
 The so called {\it 't Hooft ansatz}

\be
s(x) = F^2(x) = {48 \over { \rho^4}} \left[ 1 + 
\sum_{\mu=1}^4\left( {{x_{\mu}-x_{\mu}^0} \over {\rho}}\right)^2
\right]^{-4}, \ \ q(x) = {{Q_0} \over {8 \pi^2}} s(x)
\label{e.dens}
\ee

\noindent  describes 
an isolated instanton of size $\rho$, 
centered at $x^0$. Obviously these are scale invariant solutions. Superpositions of $N$ 
instantons {\it or}  anti-instantons also 
corresponds to (higher) minima of the action: $S = N S_0$, however
a pair instanton - anti-instanton is not a minimum and its
total action depends, among other, by the amount of ``overlap" 
$\omega = \rho_I \rho_A/ |x_I^0 - x_A^0|^2 $. 

\no Effects associated with instantons are:

{\it 1. $U_A(1)$ symmetry breaking.} This effect is succinctly described 
by the Witten - Veneziano formula\cite{wive} relating the topological susceptibility $\chi$ and
the $\eta'$ mass:

\be
\chi \equiv \lim_{V\rightarrow \infty} \frac{1}{V} 
\int dx^4 <{\rm T}(q(x)q(0))> = 
\frac{f_\pi}{2N_f} (m_\eta^2+m_{\eta'}^2-2 m_K^2)  .
\label{e.eta}
\ee

\no Here $\chi$
has to be obtained in a quenched simulation. Since the 
instantons are essentially an $SU(2)$ phenomenon, the result should not depend
strongly on $N_c$. The empirical data for the RHS of (\ref{e.eta}) give 
$\chi = (180{\rm  MeV})^4$.
Notice that the
presence of light fermions inhibits the topological susceptibility (the Dirac
operator for massless fermions has zero modes for $Q \not = 0$). 

{\it 2. Chiral symmetry breaking.} 
By inducing zero modes of the Dirac operator
instantons may control the chiral symmetry breaking, due to the 
Banks-Casher relation $<\psi \bar{\psi}> =\frac{1}{\pi} <\rho(0)>$
in which $\rho(\lambda)$ is the spectral density of the Dirac
operator.  Spontaneous symmetry breaking then results from a nonzero
density of eigenvalues at $\lambda=0$.

{\it 3. Dynamical effects.} Instantons are believed to influence the
dynamics of the intermediate distances (at the scale of $\half$ fm, say) 
and therefore the hadron properties. Their infrared effects are, however,
less certain, in particular there is an open question whether
they lead to confinement. \par

The $U_A(1)$ symmetry breaking involves only the integrated 
correlation
$\chi$. This can be tested directly in numerical simulations.
The effects in {\it 2.} and {\it 3.} depend on details of the 
instanton population. Predictions can be made with help of
models. Numerical simulations
are then asked to test the ingredients on which these models
are based. The features relevant
for our discussion are:

(i) The instantons can be in a gas,
 liquid, or crystalline phase depending 
on their density and overlap.  
The diluteness of the ensemble is expressed by 
the ``packing fraction" $f=\pi^2/2 <\rho^4> N/V$.

(ii) The size distribution controls the diluteness and the I-R properties.
In the dilute gas approximation 
the
estimation of the fluctuation determinant \cite{th76} permits to 
predict a power law for small sizes.  In case of a dense population of
instantons, 
by just integrating in a finite volume and assuming convergence of the
thermodynamic limit one can derive 
some rather general features of the size distribution \cite{gern}. 
This 
leads to a different power law behavior for the  small
size  distribution. We can write:

\be
P(\rho) \sim  \rho^p, \ \ p_{dilute} = -5 + b, 
\ \ p_{dense} = -1 + 4 (b-4)/b ,\ \ b = 11 N_c/3
\label{e.rd1}
\ee

(iii) Denoting by $N_I  (N_A)$ the number of instantons (anti-instantons)
in a configuration, and with $N = N_I + N_A$, 
we  write:

\be
c \equiv \left({\langle N^2 \rangle - \langle N \rangle^2}
\right) /{\langle N \rangle}.
 \label{e.corr}
\ee

\no  The quantity  $c$ tests the Poisson character of the $N$ - distribution.
For a poissonian distribution $c = 1$, while low
energy sum rules \cite{dia} suggest $c = {4 \over b}$.

(iv) In the ``standard" model \cite{shu}
the QCD instanton ensemble below the critical temperature is an
interacting instanton liquid with density about $N/V=$1 fm$^{-4}$.
The instanton size distribution has an infrared cut-off and
an average instanton size of $\sim 1/3$ fm.
The dynamics of the chiral phase transition is 
driven by a rearrangement 
of the instanton ensemble at $T_c$ 
-- I-A molecules are formed, with a tendency to orientation along the
euclidean time axis (this effect is due to light fermions).

\section{Method}
\begin{figure}[tb]
\vspace{4.1cm}
\includegraphics{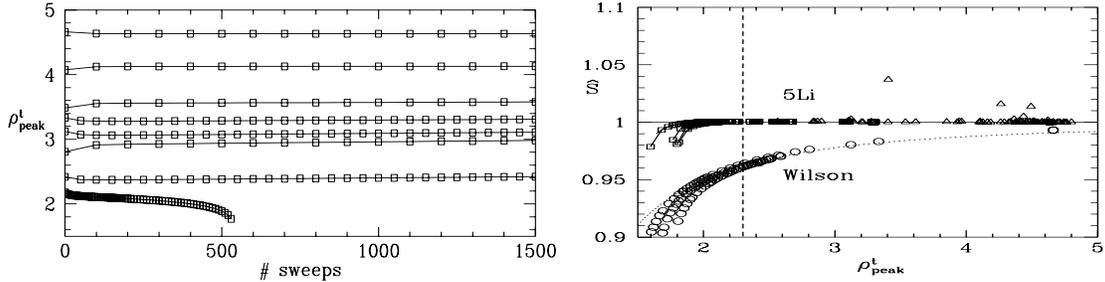}
\caption{Evolution of the size of instantons under improved cooling (left) 
and the (improved: 5Li or Wilson) action of instantons of various sizes (right). 
For the
determination of $\rho_{peak}$ see 
[6].
}
\label{f.act}
\end{figure}

Topology analyses for lattice regularized fields have the usual two sources of
systematic errors: cut-off effects and finite size effects. Short range 
topological fluctuations, at the level of the cut-off, are unphysical. They
 typically have an action lower than $S_0$ (dislocations) and
can spoil the determination 
of the susceptibility and distort the properties of the instanton ensemble.
The finite size of the lattice, on the other hand, distorts the properties of the 
ensemble at sizes beyond $l/2$ ($l$: the physical lattice size). 
To cope with this latter problem we must use a large enough lattice 
compared with
the range of distances we are interested in and check the
dependence on the boundary conditions.  
The short distance problem is
more involved. The method we have adopted for obtaining  sensible
densities is to cool the Monte Carlo configurations to 
smooth out the UV-noise, and in particular to eliminate the dislocations.
However, the usual plaquette action has, strictly speaking, no
instanton solutions since for them the action is always smaller than $S_0$
and becomes lower with decreasing size. Under Wilson cooling the instantons 
therefore shrink 
and after some time disappear. We have used instead an improved
action which has true, scale invariant instanton solutions for sizes
$\rho > \rho_0$ with a threshold $\rho_0 \simeq 2.3 a$ -- see Fig.
\ref{f.act}.  This action involves 5 Wilson loops and a similar construction
yields an improved charge operator which is practically an integer already
on rather rough configurations \cite{mnp97}.

Cooling proceeds by local minimization of the action and leads to 
the global minimum, which is determined by the topological sector $Q$.
From the point of view of our analysis it has two effects. On the one 
hand it smoothes
out the short distance fluctuations, including dislocations and makes
apparent the physical distance structure.
On the other hand it annihilates I - A pairs depending on the overlap 
between the opposite charges. 

 Since the (improved) charge operator Q stabilizes to an
integer value within 1 - 2$\%$ already after very few cooling sweeps
(5 - 10) the effects of 
type {\it 1.} above are well defined in our method without further
tuning, monitoring, etc. The susceptibility and charge distributions show good scaling
behavior which implies that they reveal a physical structure.  
The ensemble properties, on the other hand will depend, at least in part,
on the degree of cooling. This affects the discussion of 
points  {\it 2.} and {\it 3.} above and the corresponding tests for
instanton models.

\section{Analysis}

We have analyzed the $SU(2)$ and $SU(3)$ Yang-Mills theories at $T=0$. We varied the cut-off and the lattice size such as to have the same
physical volume at different discretization scales.
For the $SU(2)$ analysis we  used twisted boundary conditions \cite{mnp97}. We have studied  $QCD$ with two flavors of 
staggered fermions of 
mass $ma=0.008$ at 3 temperatures around $T_c$ on configurations 
kindly provided to us by the MILC collaboration. The lattices and the main
results are given in  Table \ref{t.all}.\par\smallskip

\begin{table}[h]
\label{t.all}
\begin{center}
\caption{Lattices and  cooling results}
\begin{tabular}{|c|c|c|c|c|} \hline
Lattice 
& Sw. 
& $\langle N \rangle$ 
& $c$ 
& $\langle N\rangle /fm^4$ 
\\
\hline 
\hline

${\bf (1)}\ SU(2), \beta=2.4$ & 20& 10.4& 0.4& 2.4 \\
\cline{2-5}
$a = 0.12 fm$      & 50& 5.0 & 0.5& 1.2 \\
\cline{2-5}
$12^4, 217 conf.$  &150& 2.45& 0.7& 0.6 \\
\cline{2-5}
$      $          &300 &  2.12& 1.0&    0.5  \\
\hline 
\hline

${\bf (2)}\ SU(2), \beta=2.6$ & 20&  63.0& 0.6& 14. \\
\cline{2-5}
$a = 0.06 fm$     & 50&  15.6& 0.65& 3.6 \\
\cline{2-5}
$24^4, 115 conf.$  &150&  6.5& 0.65& 1.5 \\
\cline{2-5}
$       $          &300&  3.8& 0.5 &  0.9  \\
\hline 
\hline

${\bf (3)}\ SU(3), \beta=5.85$ & 20& 12.10& 0.36& 1.81  \\
\cline{2-5}
$a = 0.134 fm$     & 50&  5.93& 0.5& 0.89  \\
\cline{2-5}
$12^4, 120 conf.$        &150&  2.52& 1.0 & 0.38   \\
\hline 
\hline

${\bf (4)}\ SU(3), \beta=6.0$  & 20& 21.63 & 0.5 &   3.30  \\
\cline{2-5}
$a = 0.1 fm$       & 50&  9.38& 0.5& 1.43  \\
\cline{2-5}
$16^4, 120 conf.$        &100&  5.33& 0.5& 0.81 \\
\hline 
\hline 

${\bf (5)}\ QCD, \beta=5.65$ & 20&  50.68& 0.41&1.75 \\
\cline{2-5}
$a = 0.115 fm$     & 50&  14.90& 1. & 0.51 \\
\cline{2-5}
$24^312, 31 conf.$ &100&   7.77& 0.8& 0.27 \\
\cline{2-5}
                   &150&  5.68&  1.& 0.20 \\
\hline
\hline 

${\bf (6)}\ QCD, \beta=5.725$& 20& 42.95& 0.57& 2.30 \\
\cline{2-5}
$a = 0.103 fm$     & 50&  8.86& 0.82& 0.47  \\
\cline{2-5}
$24^312, 43 conf.$    &100&  2.81& 0.98& 0.15  \\
\cline{2-5}
                   &150&  1.84& 1.03& 0.10  \\

\hline
\hline 

${\bf (7)}\ QCD, \beta=5.85$ & 20&  35.0& 0.4&    3.95  \\
\cline{2-5}
$a = 0.0855 fm$    & 50&   5.55& 0.84& 0.63  \\
\cline{2-5}
$24^312, 20 conf.$  &100&   0.50& 0.9& 0.06 \\
\hline

\end{tabular}
\end{center}
\end{table}

\begin{figure}[htb]
\vspace{4.3cm}
\includegraphics{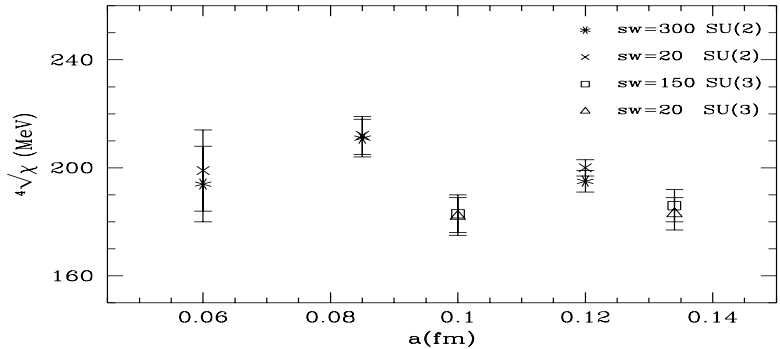}
\caption{Topological susceptibility for quenched $SU(2)$ and
$SU(3)$.}
\label{f.top}
\end{figure}

\begin{figure}[htb]
\vspace{4.5cm}
\includegraphics{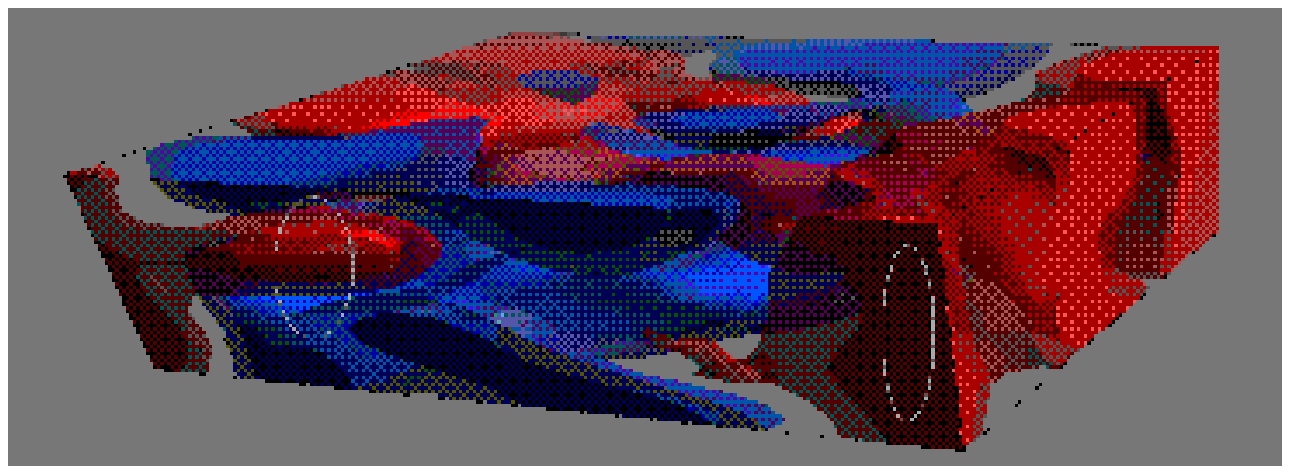}
\caption{Isosurfaces for the charge density of a typical configuration
of lattice (6), Table 1., after 20 sweeps of improved cooling; 
the two circles indicate a ``molecule" (left) and a ``caloron" (right).}
\label{f.phi}
\end{figure}

\no {\bf TOPOLOGICAL SUSCEPTIBILITY}\par\smallskip

\no As can be seen from the Fig. \ref{f.top}
the $T = 0$
topological
susceptibility is practically independent of cooling and scales correctly. $\chi^{1/4}$
shows a value of $195 - 200$ MeV for $SU(2)$ and $180-185$ MeV for $SU(3)$
with a typical error of about $5\%$, in very good agreement with experiment 
following (\ref{e.eta}) -- see also Table \ref{t.q}. 
This coincides with the results obtained using improved
operators \cite{pisa1}, underrelaxed Wilson cooling \cite{tep}, constrained 
smearing \cite{muel} and with the fermionic overlap formalism \cite{nar1} 
(where the charge density itself has been tested). The results of
\cite{boul1} are about $10\%$ higher, still compatible 
with (\ref{e.eta}) inside 2 standard deviations (see \cite{nar2} for
a possible explanation).  
We can conclude therefore that at present the $T=0$ topological susceptibility is
under control and in good agreement with the Witten-Venetiano formula.

\begin{table}[h]
\label{t.q}
\begin{center}
\caption{Average charge and topological susceptibility}
\begin{tabular}{|c|c|c|c|c|c|c|c|} 
\hline
Lattice & {\bf (1)}& {\bf (2)}& {\bf (3)}& {\bf (4)}& 
{\bf (5)}& {\bf (6)}& {\bf (7)} \\
\hline 
\hline
$\langle |Q| \rangle$& 1.6(1) & 1.5(2) & 1.8(2) & 1.7(2) &
2.0(4)& 0.9(1) & 0 \\ 
\hline
$\chi^{1/4}$ MeV & 195(4) & 194(14) & 184(6) & 182(7) & 134(10) &
102(5) & 0 \\
\hline

\end{tabular}
\end{center}
\end{table}

\begin{figure}[htb]
\vspace{5.3cm}
\includegraphics{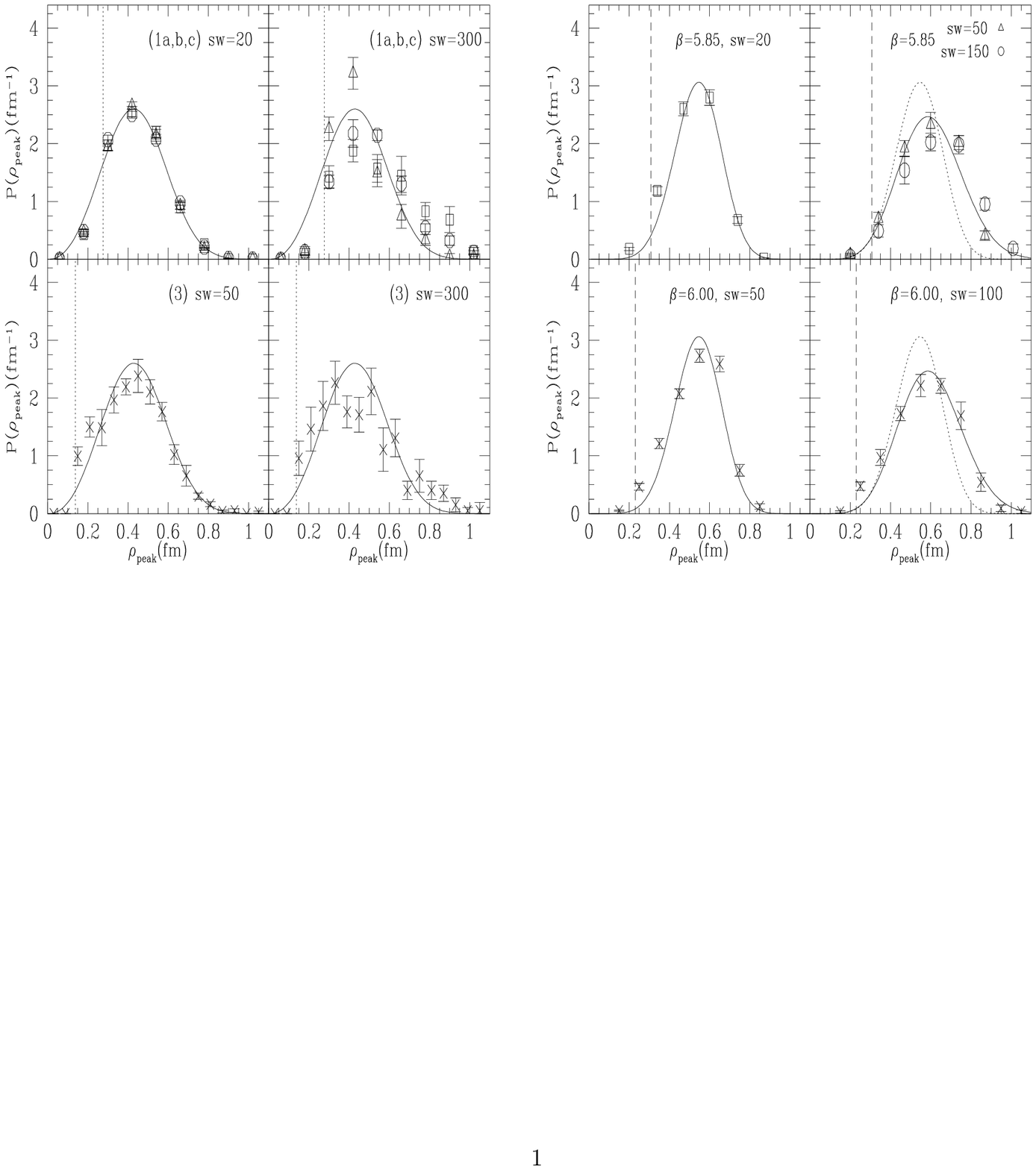}
\caption{Size distributions for quenched $SU(2)$ (left) and $SU(3)$ (right).}
\label{f.size}
\end{figure}

The finite temperature deconfining transition of the pure Yang-Mills theory 
has been shown to produce a strong drop in the susceptibility \cite{pisa1,muel}.
We have analyzed here the finite temperature QCD. Below $T_c$ we 
found as expected 
a reduction
in the susceptibility (by a factor 3), indicative of the dynamical
fermion effects. Above $T_c$ (which is at $\sim 160 MeV$ for the MILC simulation)
we found  
a pronounced drop in $\chi$, similar to   
 the quenched case. As already observed before
for other staggered fermion simulations \cite{pisa2} also 
the MILC configurations show long range correlations in $Q$ and explore few
topological sectors. The results quoted in Table \ref{t.q} are obtained 
after symmetrizing
per hand the charge distribution (since this is a symmetry of the action), the
errors given are only the statistical ones. We 
have noticed such metastabilities also in the pure $SU(3)$ simulation,
there, however, one can afford to wait until the configurations decorrelate.\par\medskip

\no {\bf THE INSTANTON ENSEMBLE}\par\smallskip

\no To observe not merely the global charge of a configuration but its structure
describable in terms of instantons and anti-instantons a certain degree of
smoothing must be attained (see Fig. \ref{f.phi} for an illustration). Using our improved cooling 
we achieve this smoothness  with a number of cooling sweeps between 
20 and 50, depending on the cut-off $a$.
We did not try to tune a rescaling law for the
amount of cooling  \cite{tep}, since for many of the 
results this was not necessary due to inherent stability properties
of our cooling.

As can be seen from the tables, when the configurations are just smooth enough for us to see instantons and anti-instantons we typically 
have a ratio $N/|Q| \simeq 6 - 10$. This ratio is reduced to  
 $N/|Q| \simeq 1 -  2$  in
the further cooling by pair annihilation,
leaving only the (anti-)instantons corresponding to 
the topological sector. 
The 
first remark is that the size distribution 
shows only little variation
during this process -- see Fig. \ref{f.size} for the quenched simulation. 
This implies that all the (anti-)instantons
which we observe from the start obey the same distribution. The average sizes
are in the region of $0.4$fm for $SU(2)$ and $0.6$fm for $SU(3)$ and
 $QCD$. The 
small size part of the distribution departs from the dilute gas power law and shows
tendency toward 
agreement with \cite{gern,gmnp}  ($p \simeq 0.8$ for $SU(2)$ and 1.5 for $SU(3)$
-- compare eq. (\ref{e.rd1})). 

The density of (anti-)instantons is high to start with.
Because of this high density and of the rather large sizes, the packing
 fraction here is also high. The overall
distribution is here 
non-poissonian and  in agreement with the low energy
sum rules, see eq. (\ref{e.corr}) and Table \ref{t.all} (while it approaches the poissonian distribution 
for instantons {\it or} anti-instantons
after long cooling). This suggests a dense ensemble 
with a spatially random
distribution biased toward small $|Q|$.
Similar results at $T=0$ have been reported in \cite{tep}.

 To the extent we could 
follow the structure described by this ensemble back to shorter 
cooling, it does not change
in character but becomes more and more obscured by the 
accumulation of short
range ripples and dislocations. 
This corroborates  with the stability of
distributions under further cooling (Fig.  \ref{f.size})
and their correct scaling behavior to suggest the
physical relevance of this ensemble. Since the short
cooling needed before we can observe the instantons and anti-instantons 
might have also annihilated some pairs in the physically relevant
size region, the figures we obtain for the density of
(anti-)instantons are  lower bounds.
\par\smallskip

For $QCD$ at high $T$ molecule formation with preferred 
orientation should be revealed by an anisotropy 
in the density - density correlations along space 
and along (euclidean) time
directions. The anisotropy  
we observe at very short cooling is compatible with zero
 at the level of two standard deviations and very low
in relative value (compared to the correlation itself) \cite{mnpl97}.  
This signal is too weak to support the scenario of \cite{shu}, 
however the quark mass might not be yet small enough to 
promote a strong effect. See also Fig. \ref{f.phi}.\par\bigskip

We conclude that we observe  
 a dense ensemble randomly distributed in space, but with a bias toward small 
$|Q|$. This ensemble is revealed after a short cooling which essentially
has filtered out dislocations and  strongly overlapping IA pairs. 
We remark as a question of principle that any separation \par

\centerline {{\it q(x)} = instantons + trivial fluctuations}\par

\no is inherently ambiguous in the case of such strongly overlapping  pairs.  It is therefore not clear to us whether
this latter structure is reasonably described as instantons and
anti-instantons or a description in terms of other kind of excitations is more
suitable for it.  \par\bigskip 
 
{\bf ACKNOWLEDGMENTS:} We wish to thank D. Diakonov, A. Di Giacomo, 
T. De Grand, A. Hasenfratz,
M. Ilgenfritz, 
M. Mueller-Preussker, G. M\"unster, R. Narayanan and T. Kov\'acs for discussions. 
MGP and IOS  thankfully acknowledge partial support from the DFG.

\end{document}